\documentclass[conference]{IEEEtran}
\usepackage[mathcal]{euscript}
\usepackage{amsmath}
\usepackage{amsthm}
\usepackage{amssymb}
\usepackage{graphicx}
\usepackage{algorithmic}
\usepackage{algorithm}
\usepackage{epsfig}
\usepackage{booktabs}% http://ctan.org/pkg/booktabs
\usepackage{colortbl}% http://ctan.org/pkg/colortbl
\usepackage{amsmath}% http://ctan.org/pkg/amsmath
\usepackage{xcolor}% http://ctan.org/pkg/xcolor
\usepackage{graphicx}% http://ctan.org/pkg/graphicx
\usepackage{cite} % aggregate refrences

\providecommand{\abs}[1]{\lvert#1\rvert}				% Absolute value
\providecommand{\abs}[1]{\lvert#1\rvert}					 % Absolute value
\providecommand{\mbf}[1]{\mathbf{#1}}						 % Math boldface
\providecommand{\mc}[1]{\mathcal{#1}}				 % Math Cal
\providecommand{\bsym}[1]{\boldsymbol{#1}}				 % Bold symbols by the bm package. Better than 																					 amsmath \pmb if the font exists.
\IEEEoverridecommandlockouts

\begin{document}

%
%\title{Coexistence Analysis between MIMO Radar and MIMO Cellular System in LoS Channel}
%
%\title{Coexistence Analysis between MIMO Radar and LTE in LoS Channel}
%
%\title{Coexistence Analysis between seaborne MIMO Radar and LTE in Littoral Areas}

\title{Coexistence Analysis between Radar and Cellular System in LoS Channel}

%\author{\begin{tabular}{c}
% Author 1, Author 2, and Author 3 
%\end{tabular}
%
%
%}
\author{\begin{tabular}{c}
 Awais Khawar, Ahmed Abdelhadi, and T. Charles Clancy \\
\{awais, aabdelhadi, tcc\}@vt.edu \\
Department of Electrical and Computer Engineering, 
Virginia Tech, VA, USA
\end{tabular}
\thanks{This work was supported by DARPA under the SSPARC program. Contract Award Number: HR0011-14-C-0027. The views, opinions, and/or findings contained in this article are those of the authors and should not be interpreted as representing the official views or policies of the Department of Defense or the U.S. Government.

Approved for Public Release, Distribution Unlimited
}

}

\maketitle

\begin{abstract}
Sharing spectrum with incumbents such as radar systems is an attractive solution for cellular operators in order to meet the ever growing bandwidth requirements and ease the spectrum crunch problem. In order to realize efficient spectrum sharing, interference mitigation techniques are required. In this letter we address techniques to mitigate MIMO radar interference at MIMO cellular base stations (BSs). We specifically look at the amount of power received at BSs when radar uses null space projection (NSP)-based interference mitigation method. NSP reduces the amount of projected power at targets that are in-close vicinity to BSs. We study this issue and show that this can be avoided if radar employs a larger transmit array. In addition, we compute the coherence time of channel between radar and BSs and show that the coherence time of channel is much larger than the pulse repetition interval of radars. Therefore, NSP-based interference mitigation techniques which depends on accurate channel state information (CSI) can be effective as the problem of CSI being outdated does not occur for most practical scenarios.

\end{abstract}

%\begin{keywords}
%Spectrum sharing, interference mitigation, MIMO radar, spectral coexistence.
%\end{keywords}

\section{Introduction}

The ubiquitous use of smart phones and tablets has resulted in tremendous growth in wireless data traffic. Current spectrum allocations are making it hard for operators to support the growth in broadband demand. In order to solve the spectrum congestion problem innovative solutions have been proposed that include design of spectrally efficient air-waves, bandwidth-rich millimeter wave communication systems, and sharing of spectrum across government agencies and commercial services. Spectrum sharing is a promising solution as it does not involve expensive and time consuming efforts to relocate incumbents for spectrum reallocation. A proposal to allow small cells to share the 3.5 GHz radar band is under consideration by the Federal Communications Commission (FCC) \cite{FCC12_SmallCells}. This is a promising initiative but considerable research is required to address the interference concerns that will arise due to the co-channel spectrum sharing of radar and small cells.

Opportunistic spectrum access schemes have been considered in past to share spectrum with traditional rotating radar systems \cite{SPC12, Bari15ciss2}. However, modern military ships are equipped with phased array radar systems that do not rotate. Furthermore, in near-future, these systems are to be replaced by MIMO radars as they promise waveform diversity, target localization, and interference mitigation capabilities that are superior than phased array radar systems \cite{LS08}. Therefore, in this paper we consider shipborne MIMO radar architecture, because of its potential near-future field deployment, for our coexistence analysis. Recently, beamforming \cite{DH13} and waveform-shaping \cite{KAC14_TDetect} based schemes have been proposed to mitigate MIMO radar interference at communication system. In this letter, we focus on the waveform-shaping approach \cite{KAC14_TDetect} that mitigates radar interference at communication systems by shaping radar waveform in a way that it falls in null space of channel between radar and communication system. NSP-based technique preserves radar mission objectives, with minor degradation \cite{KAC14_TDetect}, while allowing spectral coexistence and thus increasing available spectrum for commercial communication systems without needing to relocate radars to new frequency bands.

In this letter, we extend the previous work in \cite{KAC14_TDetect} -- which is limited to studying target detection performance of spectrum sharing MIMO radars in Rayleigh channels -- to power received at communication system, degradation in radar transmitted power, and calculation of coherence time of radar-communication system in LoS channels. The NSP technique depends on accurate CSI for effective interference mitigation. CSI is valid as long as radar's pulse repetition interval (PRI) is shorter than the channel's coherence time. CSI is acquired by radar by aiding communication systems in channel estimation, with the help of a low-power reference signal (see Sec II. \textit{I} in \cite{KAC14_TDetect}). These CSI estimates are fed back by the communication system to radar. We compute the channel coherence time for a spectrum sharing scenario in which a moving seaborne radar is sharing spectrum with an onshore communication system and show that PRI of many practical radars are much shorter than the coherence time of the channel. Thus, NSP techniques can be applied without any fear of CSI being outdated.

The rest of this letter is organized as follows. Section \ref{sec:models} briefly presents MIMO radar architecture, spectrum sharing scenario, and LoS channel model. Section \ref{sec:power} provides a discussion on interference power received at communication systems and loss in projected radar power at target. Section \ref{sec:coherence_time} computes the coherence time of channel and discusses a numerical example. Section \ref{sec:conc} concludes the letter.

\section{System Models}\label{sec:models}
In this section, we briefly introduce the fundamentals of MIMO radar, LoS channel model, and spectrum sharing scenario between radar and communication systems.

 \subsection{MIMO Radar}
 
%Previous work on spectrum sharing between radar and cellular systems focused on rotating radars \cite{} that are no longer deployed on advanced military ships. 
%MIMO radar is an active area of research and a strong contender for the replacement of legacy radar systems which utilize electronic steering and are commonly known as phased array radar systems. This is because MIMO radars promise waveform diversity, target localization, and interference mitigation capabilities that are superior than phased array radar systems \cite{LS08}. Therefore, in this paper we consider shipborne MIMO radar architecture, because of its potential near-future field deployment, for our coexistence analysis. 
We consider $M$ antenna elements and denote samples of baseband equivalent transmitted waveform as $\left\{\mbf{x}(n)\right\}^{L}_{n=1}$. In MIMO radar literature orthogonal waveforms are shown to outperform other waveforms \cite{LS08}, therefore, we design orthogonal waveforms whose signal correlation matrix is
\begin{equation}
\mbf{R}=\frac{1}{L} \sum^L_{n=1} \mbf{x}\left(n\right) {\mbf{x}}^H\left(n\right) = \mbf I
\end{equation}
where $L$ is the total number of time samples and $n$ is the time index. The signal received from a single point target at an angle $\theta$ can be written as \cite{LS08}
\begin{equation}
\mbf y(n) = \alpha \, \mbf A(\theta) \,  {\mbf {x}}(n) + \mbf w(n) \label{eqn:rxRadar}
\end{equation}
where $\alpha$ represents the complex path loss including the propagation loss and the coefficient of reflection, $\mbf w(n)$ is the white Gaussian noise, and $\mbf A (\theta)$ is the transmit-receive steering matrix defined as $\mbf A (\theta) \triangleq \mbf a(\theta) \mbf a^T(\theta)$.
%\begin{equation}
%\mbf A (\theta) \triangleq \mbf a(\theta) \mbf a^T(\theta).
%\end{equation} 
The transmit/receive steering vector $\mbf a(\theta)$  is given as
\begin{equation}
\mbf a(\theta) \triangleq \begin{bmatrix} e^{-j \omega_c \tau_{t,1}(\theta)} &e^{-j \omega_c \tau_{t,2}(\theta)} &\cdots &e^{-j \omega_c \tau_{t,{M}}(\theta)} \end{bmatrix}^T.
\label{eq:at}
\end{equation}

\subsection{Spectral Coexistence Scenario}
We consider a practical scenario where radar (incumbent) is operating in the 3550-3650 MHz band which FCC has proposed to share with commercial cellular systems on a co-primary basis \cite{FCC12_SmallCells}. Thus, opportunistic spectrum access techniques proposed -- for e.g., as in \cite{SPC12} -- are no longer valid. For the prevailing cellular standard i.e., long term evolution (LTE), 3GPP has defined Band 22 for Frequency Division Duplex (FDD) LTE (Uplink:3410-3490 MHz/Downlink:3510-3590 MHz) and  Bands 42 (3400-3600 MHz) and 43 (3600-3800 MHz) for Time Division Duplex (TDD). Since, FCC's proposed frequency range is not fully aligned with the current 3GPP band definition there is a need for a new 3GPP frequency band. So, we assume a FDD LTE deployment in which uplink is in the 3550-3650 MHz band, or radar band, and BSs get interfered by radar operations. We device a scheme for interference mitigation from radar in the uplink. Note that since downlink is assumed to be in a higher non-radar band there will be no interference to cellular users from radar systems. Without loss of generality, we consider a single cell or a BS that receives the following signal on  the uplink
\begin{equation}
{\mbf{r}}=\sum_{i=1}^{\mc K} \mbf G_i \mbf s_i + {\mbf{H}} \mbf x + \mbf{n} 
\end{equation}  
where $\mc K$ is the number of users in the cell, $\mbf G_i$ is the channel gain between the BS and the $i^{\text{th}}$ user, $\mbf H$ is the channel gain between the BS and the radar, $\mbf x$ is the interfering signal from the MIMO radar, and $\mbf n$ is the white Gaussian noise component. The interfering signal $\mbf x$ from radar can be mitigated by projecting radar waveform onto null space of interference channel such that ${\mbf{H}} \mbf x = \bsym 0$, please see \cite{KAC14_TDetect, S.SodagariDec.2012, GhorbanzadehMilcom2014, KAC14_TWS, Channel2D, Channel3D, KAC14_QPSK, A.Khawar, KAC14_Milcom, KAC14DySPANWaveform, KAC+14ICNC, SAC+15, SKA+14DySPAN, MKA_precoder} and reference therein for a discussion on this approach.

%
%
%
%
%
%
%
%So in our scenario BS and radar operating frequencies are overlapping or they are sharing spectrum where as cellular users use higher non-radar
%
%
%
%
%Assume a scenario in which a cellular system is deployed in the radar band, for example, small cells in the 3.5 GHz radar band \cite{FCC12_SmallCells}. We assume that the duplexing technique used by the cellular system is Frequency Division Duplex (FDD) in which BSs are operating in the radar band and user equipments (UEs) are operating at non-radar frequencies. The BSs send pilot/reference signals that help radar exploit channel reciprocity to obtain CSI. It is important to mention that the duplexing scheme (FDD or TDD) adopted by cellular system does not make any difference (from the point of view of radar) as long as BSs send reference signals on radar band which radar can reciprocate to find CSI. Therefore, we choose a FDD based system to focus on interference caused by the radar operation to the BSs and devise a scheme for interference mitigation.

\subsection{LoS Channel Model}
We consider a spectrum sharing scenario between a shipborne radar and a BS mounted on the top of a building or its sidewalls such that it has a LoS component with the radar. This is typical of littoral areas. Since littoral area is assumed, the area is free of reflectors or scatterers or they are very weak as compared to the LoS component and do not contribute towards the channel model. We model the LoS channel by assuming the inter-element spacing between antennas at the BS is $\Delta_N$ and at radar is $\Delta_M$, the channel matrix can be written as
\begin{equation} \label{dist_eq}
\mathbf{H}=a\sqrt{{N}M}\exp\left(-j2 \pi  \frac{d}{\lambda_c} \right) \mathbf{e}_N\left(\Omega_N\right) \mathbf{e}_M^* \left(\Omega_M\right)	
\end{equation}
where $a$ is the attenuation along the line-of-sight path which is assumed to be equal for all antenna pairs, $d$ is the distance between radar transmit antenna 1 and BS receive antenna 1, $ \lambda_c$ is the carrier wavelength, $\mathbf{e}_N$ and $\mathbf{e}_M$ are defined to be 
%\begin{eqnarray}
%\mathbf{e}_N\left(\Omega_N\right)&=&\frac{1}{\sqrt{N}} 
%\begin{bmatrix}
%1\\
%\exp\left(-j2 \pi \Delta_N \Omega_N \right)\\
%\exp\left(-j2 \pi 2 \Delta_N \Omega_N \right)\\
%\vdots \\
%\exp\left(-j2 \pi \left(N-1\right) \Delta_N \Omega_N \right)\\
%\end{bmatrix}\\
%\mathbf{e}_M \left(\Omega_M\right)&=&\frac{1}{\sqrt{M}} 
%\begin{bmatrix}
%1\\
%\exp\left(-j2 \pi \Delta_M \Omega_M \right)\\
%\exp\left(-j2 \pi 2 \Delta_M \Omega_M \right)\\
%\vdots \\
%\exp\left(-j2 \pi \left(M-1\right) \Delta_M \Omega_M \right)\\
%\end{bmatrix}
%\end{eqnarray}
\begin{eqnarray}
\mathbf{e}_l\left(\Omega_l\right)&=&\frac{1}{\sqrt{l}} 
\begin{bmatrix}
1\\
\exp\left(-j2 \pi \Delta_l \Omega_l \right)\\
\exp\left(-j2 \pi 2 \Delta_l \Omega_l \right)\\
\vdots \\
\exp\left(-j2 \pi \left(l-1\right) \Delta_l \Omega_l \right)
\end{bmatrix}
\end{eqnarray}
where $l=\{N,M\}$, $\Omega_M \triangleq \cos\phi_M$ and $\Omega_N \triangleq \cos\phi_N$ are the angles of incidence of the line-of-sight path on the radar and BS antenna arrays, respectively.

\section{Received Power Analysis}\label{sec:power}

In this section, we look at the amount of power received at BSs and the target. We are interested in knowing about the power received at BSs and target for effective interference mitigation and target detection purposes, respectively. The gain of the radar transmit array in a direction $\theta$ when the beam is steered digitally to a direction $\theta_D$ is given by \cite{LS08}
\begin{equation} \label{bptx}
G (\theta,\theta_D)=\Gamma \dfrac{\abs{\mbf a^H(\theta) \mbf R^T \mbf a(\theta_D)}^2}{\mbf a^H(\theta_D) \mbf R^T \mbf a(\theta_D)}
\end{equation}
where $\Gamma$ is the normalization constant. We are interested in placing nulls or having minimum gain towards the direction of BSs, by using NSP-based interference mitigation scheme, and maximum gain in the direction of target. In the following sections we cover both the scenarios in detail along with examples.

\subsection{Power Received at Cellular System}
\begin{figure}
\centering
\includegraphics[width=3in]{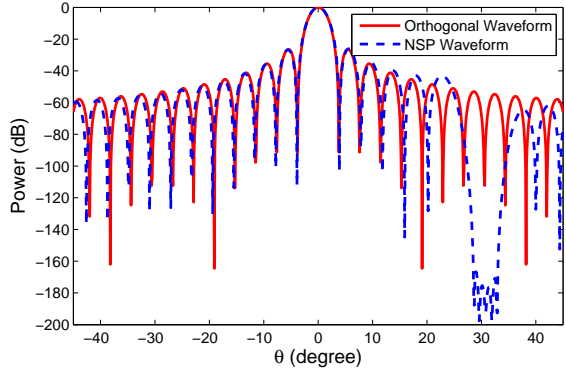}
\caption{Radar's transmit beampattern. NSP-based interference mitigation scheme places accurate and deep nulls in the location of BSs ($30^\circ$ to $35^\circ$) for effective interference mitigation.}
\label{fig:rx_power}
\end{figure}

In this section we study the received power at locations nulled by radar system using null space projection algorithm. These nulled locations are occupied by cellular BSs and are subject to interference protection from radar system. In Figure \ref{fig:rx_power} we show this scenario when the target is located at $0^\circ$ and BSs are located at an azimuth of $30^\circ$ to $35^\circ$. Note that the received power at BS locations is much below the power projected at target and other azimuthal locations. The NSP places accurate and deep nulls at locations that are occupied by cellular BSs. The received power level is much below the noise floor of most practical BSs. For example, LTE eNode B has a noise floor of -120 dBm (-150 dB) \cite{NTIA13}.  Thus, radar interference can be effectively mitigated by using the proposed NSP-based algorithm.

\subsection{Reduction in Power Projected at Target}

\begin{figure}[!]
\centering
\includegraphics[width=3in]{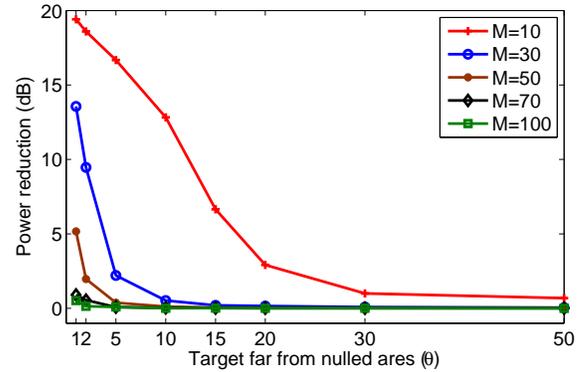}
\caption{Power reduction in mainlobe due to NSP as a function of angular separation between communication system and target.}
\label{pow_red}
\end{figure}

In this section, we evaluate the power reduction in the main beam due to NSP. As an example, we consider the case where BSs are present at an azimuth angle of $20^\circ$ to $25^\circ$ and the target first appears at an azimuth angle of $26^\circ$, with respect to (w.r.t.) radar. Thus, there is a angular separation of $1^\circ$ between the communication systems and the target. We further increase this angular separation to $2^\circ, 5^\circ, 10^\circ, 15^\circ, 20^\circ, 30^\circ$, and $50^\circ$ and study reduction in mainlobe power. So the reduction in mainlobe power is presented as a function of angular difference between the communication systems and the target. Moreover, we analyze this power reduction by employing $M=10, 30, 50, 70$ and 100 antenna elements at MIMO radar while the antennas elements at communication system are fixed at $N=5$. The results are reported in Figure \ref{pow_red}. It can be noted that for a small radar array and a target immediately next to the nulled zone, i.e when $N=10$ and target is at $1^\circ$ relative to BSs, the loss in projected power is much more severe than all the other cases. As the target moves away from the communication systems the loss in power projected becomes smaller and smaller. Moreover, when the radar employs a larger antenna array, for example with 70 or 100 elements, the power reduction in mainlobe, due to NSP, is negligible. 

\section{Coherence Time Of A Shipborne Radar} \label{sec:coherence_time}
Null-space based projection scheme requires CSI estimation. However, it has not been investigated that for how long the CSI is valid and after what time period the CSI becomes outdated. Specifically, knowledge about the coherence time of the channel between shipborne radar and cellular system is not available. In this letter, we investigate this issue and derive coherence time of channel between shipborne radar and stationary communication system. The movement of a ship, and hence radar, is affected by factors such as wind speed, length of time the wind blows, distance of open water over which the wind blows (i.e. fetch), see Table \ref{fig:V_bob}; because these factors gives rise to waves which affect the motion of a ship. Thus, this work is different from the classical work done on finding coherence time of channel between BS and static/mobile user \cite{And05} in a way that in addition to ship's horizontal motion (speed) we consider ship's vertical motion (bob) induced by sea.

\begin{figure}[!]
\centering
\includegraphics[width=3in]{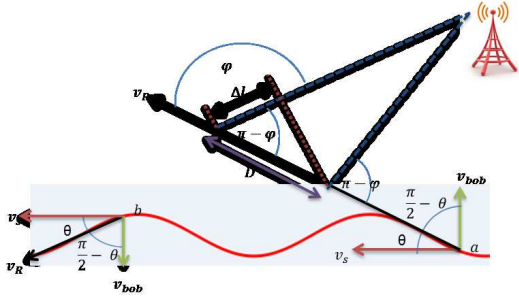}
\DeclareGraphicsExtensions{png}
\caption{Coherence time analysis of a moving shipborne radar. }
\label{coh_time}
\end{figure}

  Consider a ship-borne radar, as shown in Figure \ref{coh_time}, moving with a constant horizontal velocity $v_s$ to point $a$.  Rough seas give rise to waves that are steep, where steepness of a wave is the ratio of wave height to the length of wave, which in turn introduces bobbing velocity $v_{bob}$. Assume the ship is moving at speed $v_R$ which is the resultant of $v_s$ and $v_{bob}$. So, $v_R$ is given by,
\begin{eqnarray}
v_R&=& v_s \cos\left(\theta\right) +	v_{bob} \cos{\left({\frac{\pi}{2}}-{\theta}\right)}
%\\
%&=& v_s \cos\left(\theta\right) +	v_{bob} \sin{\left(\theta\right)}
\end{eqnarray}
where $\theta={\tan}^{-1}(v_{bob}/{v_s})$ and using values in Table \ref{fig:V_bob}, $v_{bob}$ is given by
\begin{equation}
v_{bob} = 2\: v_s \: \: \frac{\text{Height of wave}}{\text{Length of wave}}
\end{equation}
At speed $v_R$, the ship-borne radar moves along a path segment $D$ while it illuminates its search space which also contains a remote communication system. The difference in path length traveled by the waves between the two points along $D$ to the communication system can be written as:
\begin{eqnarray}
\Delta l&=& D \cos\left(\pi-\varphi\right)=v_R \Delta t \cos\left(\varphi\right)
\end{eqnarray}
where $\Delta t$ is the time required for the ship to travel the path segment $D$. Since, the communication system is assumed to be far away, $\phi$ is assumed to be the same at the two ends of $D$. The phase change of the signal received at the communication system corresponding to this difference in path lengths is therefore
\begin{equation}
	\Delta \alpha=\frac{2 \pi \Delta l}{\lambda}=\frac{2 \pi v_R \Delta t}{\lambda} \cos\left(\phi\right).
\end{equation}
So, the apparent change in frequency, or Doppler shift, is given by
\begin{equation}
	f_d=\frac{1}{2 \pi} \cdot \frac{\Delta \alpha}{\Delta t}=\frac{v_R}{\lambda} \cos\left(\phi\right).
\end{equation}
Consequently, coherence time, which is the time domain dual of Doppler spread, is given by \cite{And05},
%\begin{equation}\label{dop_1}
%T_c\approx \frac{1}{f_m}=\frac{\lambda}{v_R}	
%\end{equation}
%where $f_m$ is the maximum Doppler shift. If the coherence time is defined to be the time over which the time correlation is above 0.5, then the coherence time is approximately 
%\begin{equation}\label{time_corr}
%	T_c\approx \frac{9}{16 \pi f_m}=\frac{9 \lambda}{16 \pi v_R}
%\end{equation}
%In practice, equation (\ref{dop_1}) is too loose while equation (\ref{time_corr}) is too restrictive. So, a rule of thumb in modern digital communication is to take the geometric mean of both equations,
\begin{equation}\label{dop_2}
	T_c=\sqrt{\frac{9}{16 \pi f_m^2}} =\frac{0.423 \lambda}{v_R}
\end{equation}
where $f_m$ is the maximum Doppler shift.

\begin{figure}[!]
\centering
\includegraphics[width=3in]{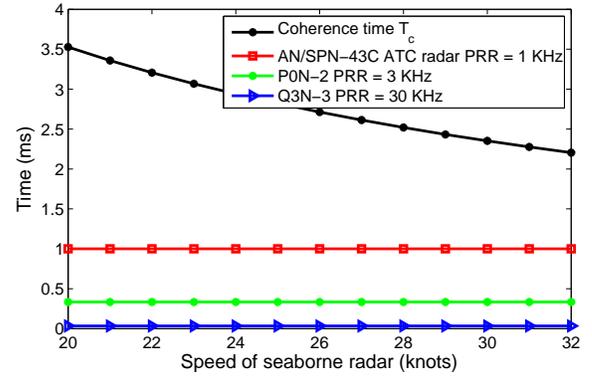}
\caption{The problem of CSI being outdated for the application of NSP does not occur as the coherence time of radar-BS channel is much larger than PRI of most practical radars. }
\label{fig:coherence}
\end{figure}

\noindent
\textbf{Example: Coherence time analysis of a moving ship-borne radar and a static BS --}
In this example, we study the relationship between coherence time of channel and NSP and seek to answer the question about the applicability of NSP for a moving radar. Consider an AN/SPN-43C air traffic control (ATC) radar, used by navy in the 3.5 GHz band, with a pulse repetition rate (PRR) of 1000 Hz or pulse repetition interval (PRI) of 1 millisecond (ms) \cite{NTIA10}. Such radars are mounted on ships that typically move with a top speed of 32 knots. Also consider radars that transmit fixed-frequency carrier wave pulse modulated waveform and swept-frequency carrier wave pulse modulated waveform. These are referred to as P0N and Q3N, respectively, in the National Telecommunications and Information Administration (NTIA) report \cite{NTIA13}. Usually, PRI, speed, and other parameters of a military radar or ship are confidential. Therefore, we use the sample information provided by NTIA in its assessment reports \cite{NTIA12, NTIA13}. Using this information the coherence time of channel is calculated and shown in Figure \ref{fig:coherence} for various operating conditions by varying ship's speed and considering different values of wind speed, wave height, wave length, for a 200 nautical mile fetch of wave. These calculations are reported in Table \ref{fig:V_bob}. It can be observed that since the PRI of radar is much smaller than the coherence time, therefore, NSP will be working perfectly even with a moving shipborne radar.

\begin{table}
\centering
	\includegraphics[width=2.5in]{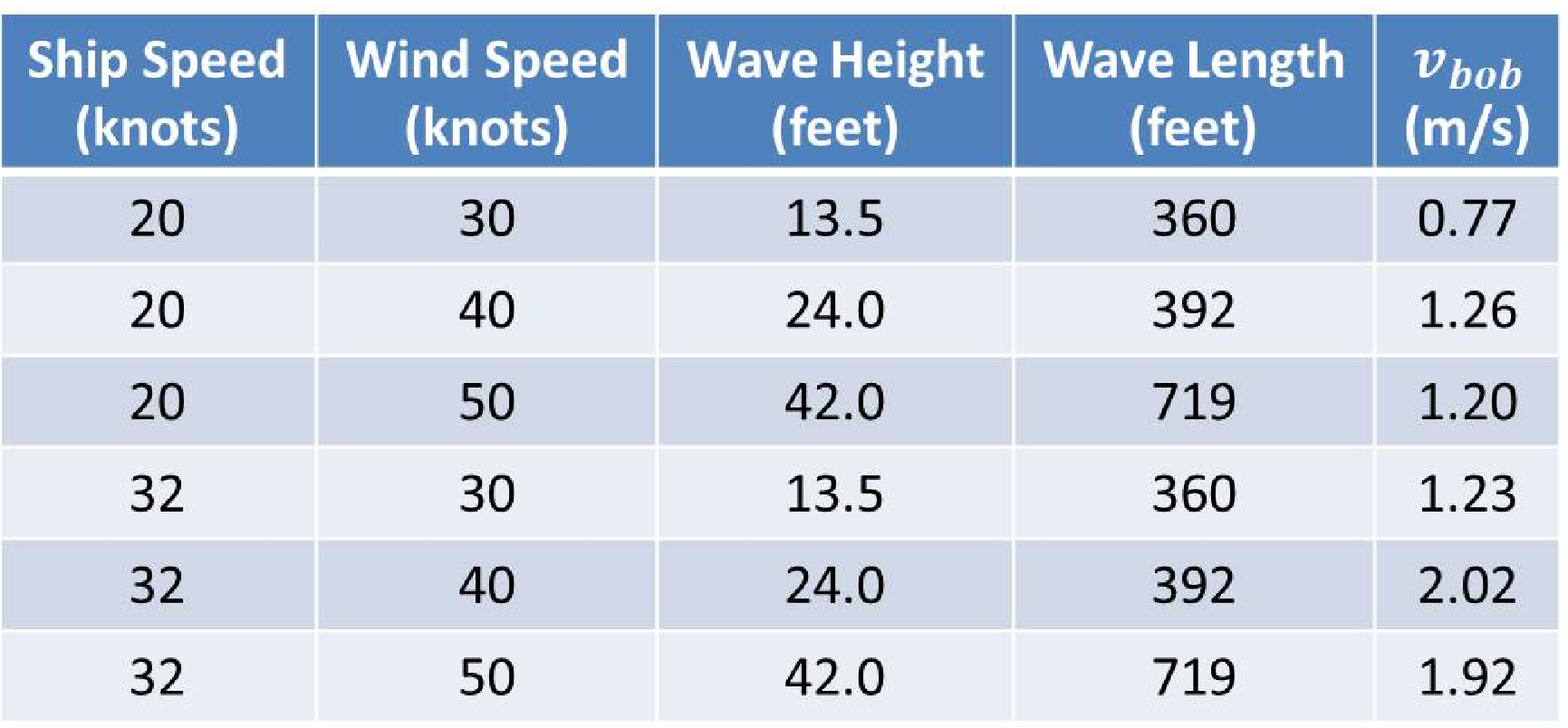} 
\caption{Values of $v_{bob}$ for various operating speeds and environmental conditions. }
%Units: meters/sec (m/s); nautical miles (nm); kilometers (km); feet (ft); meters (m).}
		\label{fig:V_bob}
\end{table}

%can be calculated by considering the steepness of wave

\section{Conclusion}\label{sec:conc}

In this letter, we evaluated a spectrum sharing scenario between seaborne radar and an onshore cellular systems. We showed that the nulls placed in the direction of BSs resulted in received power well below the noise floor of commercial BSs thus mitigating radar interference. However, the interference mitigation scheme employed resulted in loss of radar's projected power at targets that were immediately next to BS locations in the azimuth. We showed that this problem can be compensated by using a large radar antenna array. In addition, we showed that the coherence time of radar-BS channel was large enough for the application of NSP-based interference mitigation scheme which relied on CSI estimation. Thus, the issue of CSI being outdated did not arise in the radar-cellular system spectrum sharing scenario.

\bibliographystyle{ieeetr}
\bibliography{IEEEabrv,coexist2Danalysis}
\end{document}